\documentclass{aa}

\usepackage{graphicx}
\usepackage{txfonts}

\begin{document}

\title{On the nature of buckling instability in galactic bars}

\author{Ewa L. {\L}okas
}

\institute{Nicolaus Copernicus Astronomical Center, Polish Academy of Sciences,
Bartycka 18, 00-716 Warsaw, Poland\\
\email{lokas@camk.edu.pl}}

\abstract{
Many strong simulated galactic bars experience buckling instability, which manifests itself as a vertical distortion
out of the disk plane, and later dissipates. Using a simulation of an isolated Milky Way-like galaxy, I demonstrate
that the phenomenon can be divided into two distinct phases. In the first one, the distortion grows and its pattern
speed remains equal to the pattern speed of the bar, so that the distortion remains stationary in the reference frame
of the bar. The growth can be described with the mechanism of a driven harmonic oscillator with time-dependent force,
which decreases the vertical frequencies of the stars. At the end of this phase, most bar-supporting orbits have
banana-like shapes with a resonant vertical-to-horizontal frequency ratio close to two. The increase of amplitudes of
vertical oscillations leads to the decrease of the amplitudes of horizontal oscillations and the shrinking of the bar.
The mass redistribution causes the harmonic oscillators to respond adiabatically and increase the horizontal
frequencies. In the following second phase of buckling, the pattern speed of the distortion increases -- reaching one
third of the circular frequency -- but it decreases with radius. The distortion propagates as a kinematic bending wave
and winds up, leaving behind a pronounced boxy/peanut shape. The increased horizontal frequencies cause the
weakening of the bar and the transformation of banana-like orbits into pretzel-like ones, except in the outer part of
the bar, where the banana-like orbits and the distortion survive. The results strongly suggest that the buckling of
galactic bars is not related to the fire-hose instability, but it can be fully explained by the mechanism of vertical
resonance creating the distortion that later winds up.}


\maketitle

\section{Introduction}

\nolinenumbers

Many, perhaps even a majority, of late-type galaxies in the Universe possess an elongated structure in their inner
parts, which is known as a bar. For about half a century we have known that galactic disks are generally unstable and
form bars spontaneously \citep{Hohl1971, Ostriker1973}, and this mechanism has been adopted as the main channel for the
formation of bars. Another possibility, identified later on \citep{Gerin1990, Noguchi1996}, involves interactions with
other galaxies or bigger structures such as groups and clusters. Tidal forces naturally present during such encounters
lead to the elongation of the stellar component and the formation of the bar if the circumstances of the interactions
are favorable in terms of a small enough pericenter distance and prograde configuration \citep{Lokas2014, Lokas2016,
Lokas2018, Peschken2019}. The formation and evolution of bars has been studied for decades using $N$-body simulations,
and later on refinements such as the additional gas component and processes such as star formation and feedback. The
dark-matter halo, originally introduced as a way to moderate the bar instability, has been shown to play a crucial role
in bar formation due to its importance in the transfer of angular momentum \citep{Athanassoula2003}.

The simulated bars, after a period of rapid growth, tend to become thicker later on, forming a bulge-like component
with a boxy/peanut (BP) shape \citep{Combes1981, Athanassoula2005}. This shape has been observed in many spiral
galaxies viewed edge-on \citep{Kruk2019}, including our own Milky Way \citep{Weiland1994, Ciambur2017}. The thickening
is often preceded by a short episode of vertical distortion or bending of the bar out of the disk plane; it was first
described by \citet{Friedli1990}, \citet{Pfenniger1991}, and \citet{Raha1991} and is often referred to as buckling
instability. However, it should be noted that the BP shape can also form without buckling \citep{Combes1990,
Pfenniger1991, Sellwood2020, Li2023}.

In addition to creating the BP shape, buckling also weakens the bar, but generally does not destroy it, and in most
configurations it starts to regrow after the event. The instability may occur more than once in the lifetime of a bar,
with the second episode usually lasting longer and happening in the outer parts \citep{Martinez2006,
Kataria2024}. It has also been shown in idealized simulations that the presence of the gaseous component tends to
suppress buckling \citep{Debattista2006, Berentzen2007, Villa2010, Athanassoula2013, Lokas2020}. However, more
realistic configurations, modeled in cosmological simulations, do not seem to support this conclusion
\citep{Anderson2024}. There is also no obvious correlation between the presence of a BP bulge, probably originating
from buckling, and the global gas content in observed galaxies \citep{Erwin2017}.

\begin{figure*}
\centering
\includegraphics[width=18.5cm]{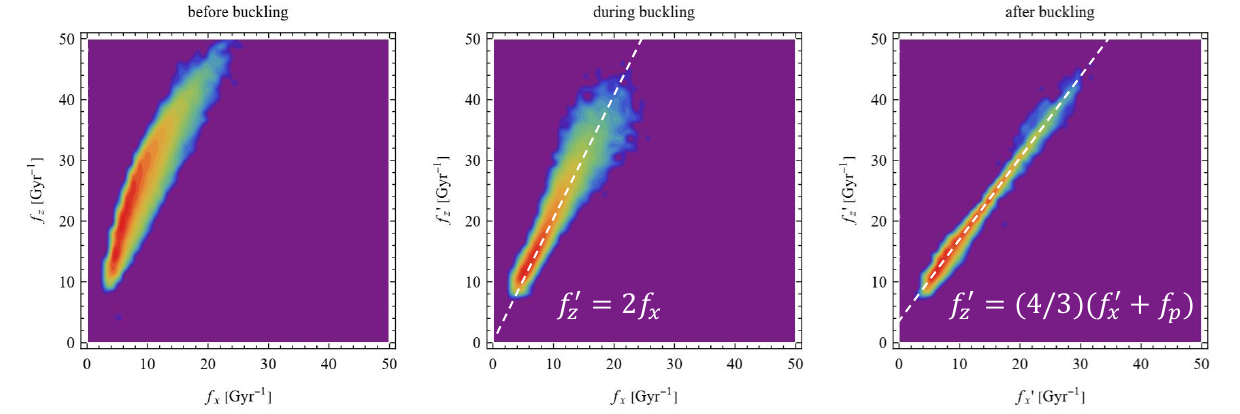}\\
\caption{
2D histograms of distribution of frequencies along the bar ($f_x$) and along the vertical direction ($f_z$) for
bar-supporting stellar orbits that underwent buckling. The three panels from the left to the right show the
distributions before buckling, during buckling, and after buckling, respectively. The white dashed lines and formulae
indicate the approximate linear relations between the frequencies at the vertical resonance (middle panel) and after
buckling (right panel).}
\label{freqzx}
\end{figure*}

Two hypotheses were proposed concerning the nature of buckling instability in galactic bars. \citet{Combes1990} and
\citet{Pfenniger1991} suggested that the formation of the BP shape results from the trapping of typical orbits of the
bar at vertical resonances. An alternative hypothesis for the origin of buckling involves the ratio of the
vertical-to-horizontal velocity dispersion of the stars in the bar and relates it to the fire-hose instability known
from plasma physics \citep{Toomre1966, Raha1991, Merritt1994}. In this picture, the instability is supposed to be
triggered by a sufficiently low ratio of these dispersions (on the order of 0.3), which is characteristic of strong
bars. Recently, the hypothesis based on vertical resonance seems to be getting more support from numerical studies
\citep{Li2023, McClure2025}.

In this work, I revisited the issue of the origin and the nature of buckling instability in galactic bars using the
insights from recently published works. The next section describes the recent progress that has been made in the field
and how it can be used to build a convincing model of buckling instability. Sections 3 and 4 describe two phases of
buckling into which the phenomenon can be divided, and Section 5 shows how the evolution of the orbital structure of the
bar can be understood within this scenario. The discussion follows in Section~6.

\section{Insights from recent studies}

In \citet[hereafter L19]{Lokas2019}, I studied buckling instability using a collisionless simulation of a Milky
Way-like galaxy evolving in isolation. The galaxy $N$-body model was initiated with two components: a spherical
dark-matter halo and an exponential disk, each containing $10^6$ particles. The dark-matter halo had a
Navarro-Frenk-White \citep{Navarro1997} profile with a virial mass of $M_{\rm H} = 10^{12}$ M$_{\odot}$ and
concentration of $c=25,$ while the exponential disk had a mass of $M_{\rm D} = 4.5 \times 10^{10}$ M$_{\odot}$,
scale-length of $R_{\rm D} = 3$ kpc, and thickness of $z_{\rm D} = 0.42$ kpc. The evolution of the galaxy was followed
for 10 Gyr (with outputs saved every 0.05 Gyr) and revealed the bar to be forming rather slowly, reaching a maximum
strength around $t = 4.3$ Gyr and the first buckling episode taking place at $t = 4.5$ Gyr. The reader is encouraged to
browse this earlier paper (L19), because the results presented below refer to and build on previous findings.

As discussed in L19, in order to study the first buckling event in more detail, the simulation was rerun between $t =
3.5$ and $t = 5.5$ Gyr, saving outputs every 0.001 Gyr (2001 outputs in total), which allowed me to follow stellar
orbits in sufficient detail to measure their properties. The properties of the orbits were estimated in the Cartesian
reference frame of the bar with $x$, $y$, and $z$ aligned with the major, intermediate, and minor axes of the bar.
Then, I calculated the amplitudes of the stellar oscillations along the three axes, $a_x$, $a_y$, and $a_z$, and the
frequencies of the orbits, $f_x$, $f_y$, and $f_z$, using a discrete Fourier transform in two periods of 0.7 Gyr before
and after buckling; that is, between 3.5 and 4.2 Gyr and between 4.8 and 5.5 Gyr, respectively. In the following, the
quantities measured before buckling are denoted by symbols without primes, while those measured after buckling are
denoted by the same symbols with primes; hence, $f_x$ and $f_x'$ will correspond to $f_{x, {\rm before}}$ and $f_{x,
{\rm after}}$ used in L19, for example. The intermediate period, between 4.2 and 4.8 Gyr, was left out because at these
times the orbits vary strongly, and the spectral analysis of stellar orbits is not reliable. It should be emphasized
that the measurements of orbital properties were done ``in vivo'', that is, in the live evolving bar rather than in a
frozen potential, as is often done in the literature. Selecting bar-supporting orbits with $a_x < 7$ kpc (smaller than
the length of the bar), $a_y < 0.7 a_x$ (elongated along the bar), and those that underwent buckling -- that is, those
that increased their $a_z$ amplitude during the evolution -- gave a sample of almost $2 \times 10^5$ orbits that were
studied further.

The most important finding of that study was the discovery of a very tight linear relation between the orbital
frequencies of the stellar orbits after buckling, which is very well fit (with the correlation coefficient
of 0.99) by the relation $f_z' = 1.35 f_x' + 3.39$ and can be approximated by the formula
\begin{equation}        \label{relationfxfz}
        f_z' = (4/3) (f_x' + f_{\rm p}).
\end{equation}
The parameter $f_{\rm p}$ was identified numerically as the pattern speed of the bar, which did not vary strongly
during buckling, and for this particular simulation it was equal to $f_{\rm p} = 2.5$ Gyr$^{-1}$. It might be useful to
note that in the more commonly used notation, with the circular frequency $\Omega' = f_x' + f_{\rm p}$ and the vertical
frequency $\nu' = f_z'$, this relation translates to an even simpler formula: $3 \nu' = 4 \Omega'$. The relation, in
terms of the 2D histogram of the distribution of the frequencies and the line given by Equation~(\ref{relationfxfz}),
is shown in the right panel of Fig.~\ref{freqzx}. Its origin is explored further below and in Section~4, where its
connection to the pattern speed of the vertical distortion is described.

The existence of this very tight relation between the vertical and horizontal frequencies after buckling was later
confirmed by \citet{Sellwood2020}, \citet{Li2023}, and most recently by \citet{McClure2025} (for the case without the
initial bulge), although they did not acknowledge the role of the pattern speed of the bar. However, using their
published results concerning the appearance of the relation and the pattern speed of their simulated bars, one can
easily demonstrate that the relation~(\ref{relationfxfz}) is also obeyed by their results. It is especially noteworthy
given that the pattern speeds of the bars are different in all these simulations. This confirms that the presence of
the pattern speed in this relation is not just a numerical coincidence, it has a deeper meaning.

Comparing this relation with the much wider distribution of frequencies before buckling (left panel of
Fig.~\ref{freqzx}), one can see that after buckling the relation becomes much tighter, and its slope changes
significantly. While before buckling the distribution barely touches the $f_z = 2 f_x$ region (from above), after
buckling all the stars are well below it. This strongly suggests that they must have crossed the vertical 2:1 resonance
-- that is, the $f_z = 2 f_x$ line -- at a certain time. It is difficult to verify this using the spectral analysis of
stellar orbits because the latter has to be performed on time spans containing many oscillations. Instead,
\citet{Li2023} recently managed to determine the instantaneous frequencies of stellar orbits in their simulation of a
buckling bar by measuring  their periods directly. They confirmed that during buckling, the majority of stellar orbits
in the bar indeed cross the 2:1 vertical resonance and thus provide strong arguments in favor of this mechanism as that
responsible for triggering the vertical distortion.

\begin{figure}
\centering
\includegraphics[width=8.9cm]{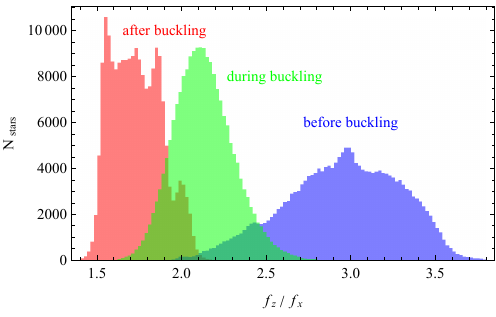}\\
\caption{
Histograms of vertical-to-horizontal frequency ratios at different stages of evolution. The blue, green, and red
distributions show the values of $f_z/f_x > 2$ (before buckling), $f_z'/f_x \sim 2$ (during buckling), and
$f_z'/f_x' \le 2$ (after buckling), respectively.}
\label{frequencyratios}
\end{figure}

\begin{figure*}[!ht]
\centering
\includegraphics[width=5cm]{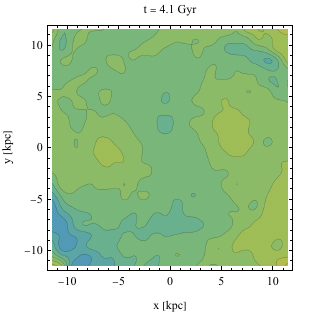}
\includegraphics[width=5cm]{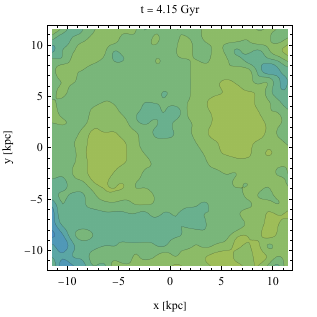}
\includegraphics[width=5cm]{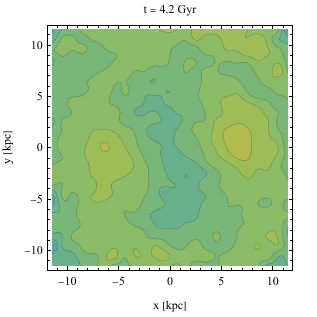}\\
\vspace{0.15cm}
\includegraphics[width=5cm]{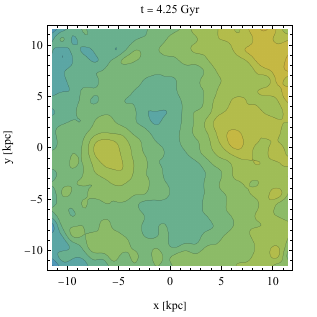}
\includegraphics[width=5cm]{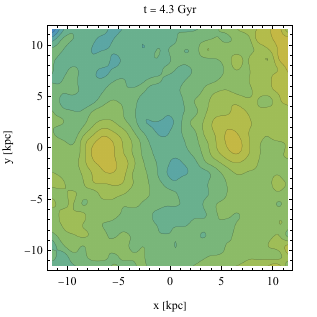}
\includegraphics[width=5cm]{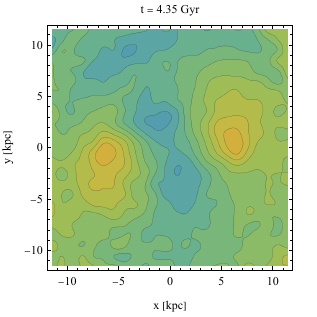}\\
\vspace{0.15cm}
\includegraphics[width=5cm]{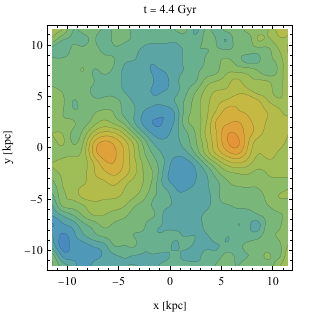}
\includegraphics[width=5cm]{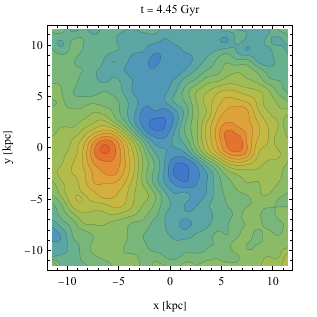}
\includegraphics[width=5cm]{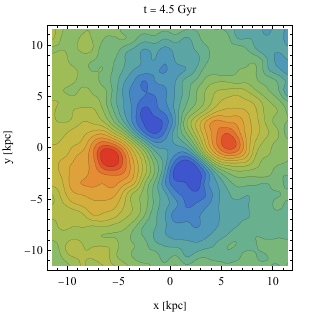}\\
\vspace{0.15cm}
\includegraphics[width=5cm]{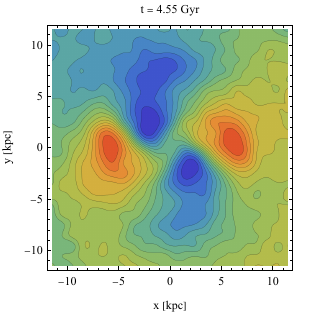}
\includegraphics[width=5cm]{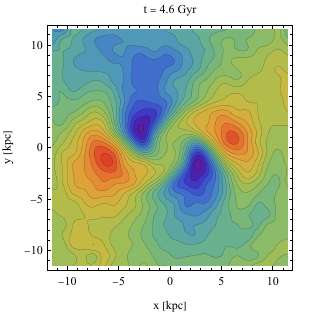}
\includegraphics[width=5cm]{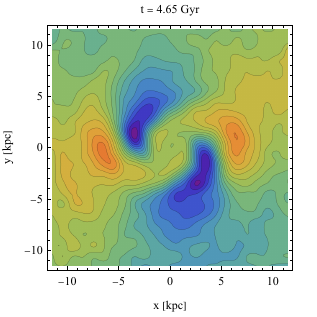}\\
\vspace{0.15cm}
\hspace{0.25cm}
\includegraphics[width=4.06cm]{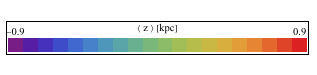}\\
\caption{
Average position of stars along vertical direction at times $t = 4.1$ to 4.65 Gyr from the start of the simulation.
Redder colors indicate the distortion above the disk plane, and the bluer ones show the distortion below it. The maps
show the measurements in the reference frame of the bar, which is oriented along the $x$-axis.}
\label{distortion1}
\end{figure*}

\begin{figure*}
\centering
\includegraphics[width=5.cm]{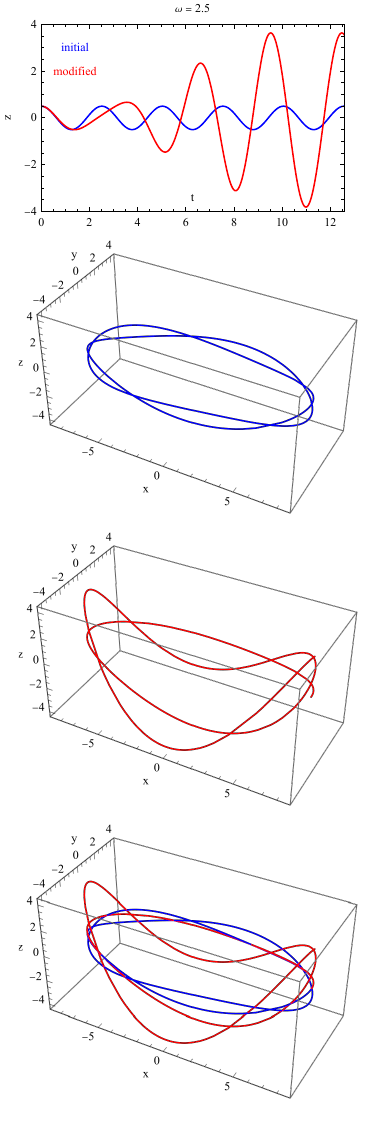}
\hspace{0.2cm}
\includegraphics[width=5.cm]{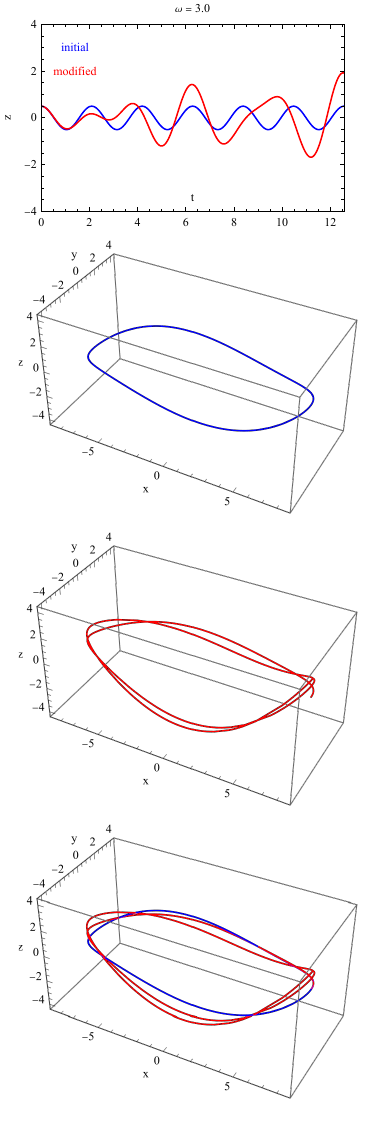}
\hspace{0.2cm}
\includegraphics[width=5.cm]{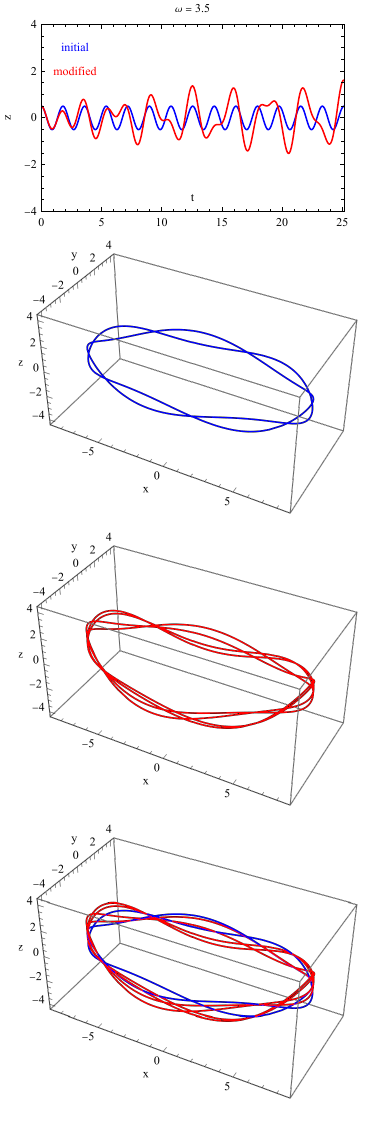}\\
\caption{
Reaching resonance. The three columns show three examples of the evolution of stellar orbits under the
growing force due to the distortion for different values of the initial frequency: $\omega = 2.5$, 3, and 3.5. The
solutions to Equation (\ref{drivenoscillator}) were obtained with the force parameters $F_0=10$ and $\alpha = 0.1$ in
all cases. The blue lines correspond to the initial oscillations and orbits, while the red lines show the modified ones.
The panels of the upper row show the variation of the vertical position of the star along the orbit, while the next rows
show the full orbits combining the motion along $z$ with an ellipse in the $xy$ plane without (blue, second row) and with
(red, third row) the action of the force and the two combined (lower row).
}
\label{resonance}
\end{figure*}

\begin{figure}
\centering
\includegraphics[width=8.5cm]{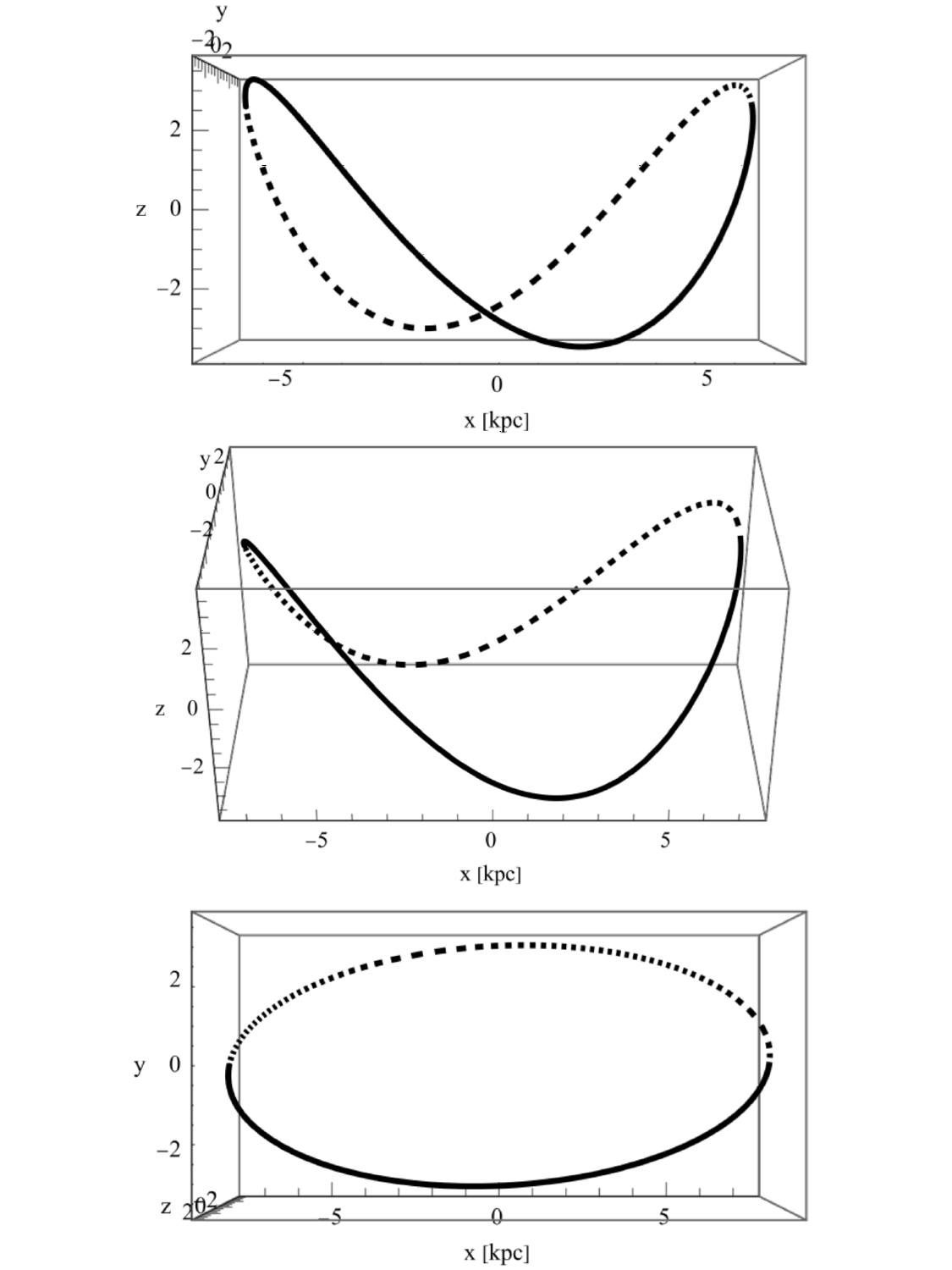}\\
\caption{
Average orbit of stars at time of vertical resonance, in first stage of buckling, viewed along three
directions: edge-on (upper panel), intermediate (middle panel), and face-on (lower panel). The dashed line indicates the
more distant part of the orbit in the edge-on view.}
\label{resorbit}
\end{figure}

\begin{figure}
\centering
\vspace{-0.5cm}
\includegraphics[width=8cm]{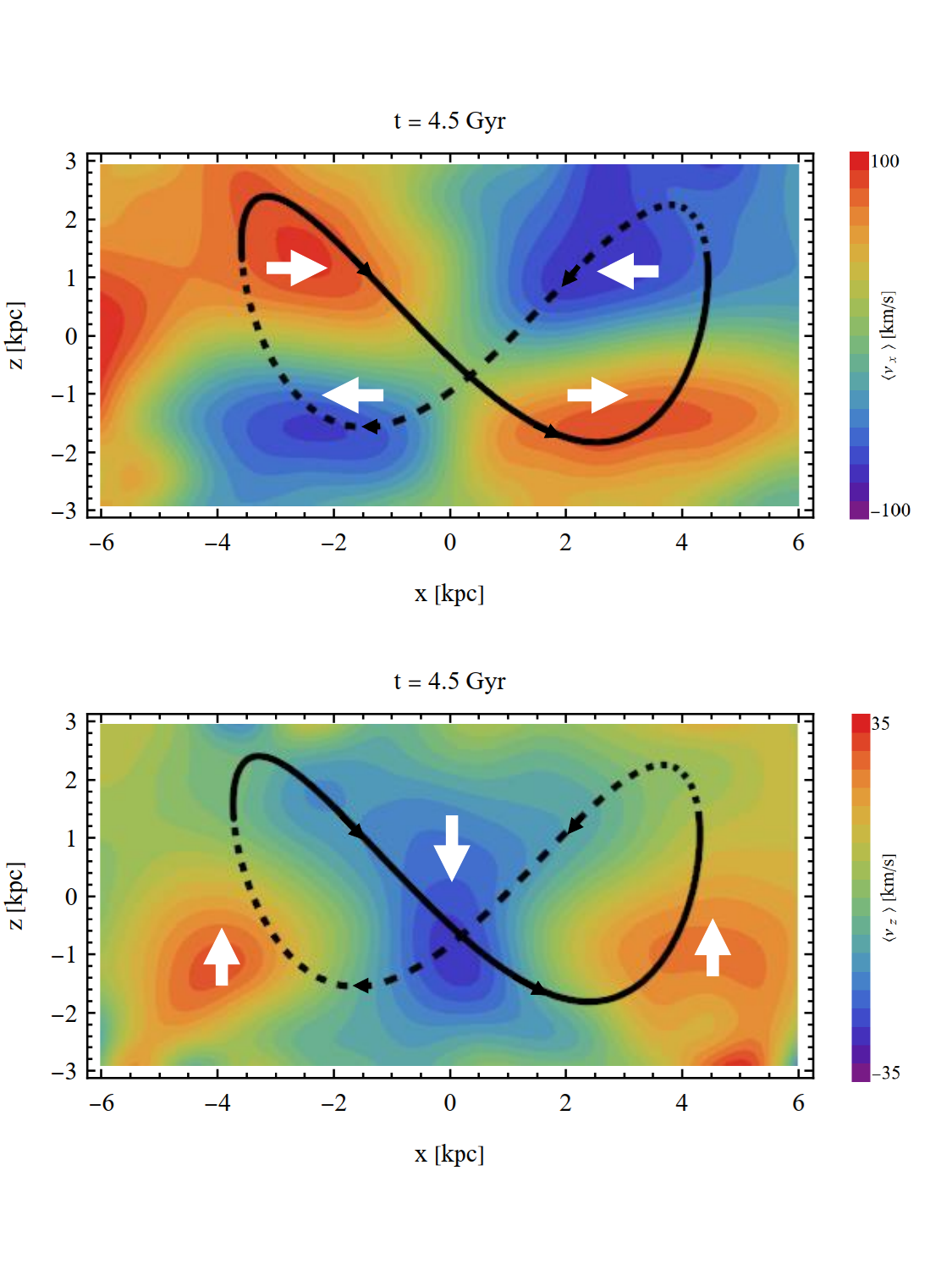}\\
\vspace{-0.7cm}
\caption{
Mean velocity of stars along $x$ (upper panel) and along $z$ (lower panel) in the bar at the time of maximum
distortion, $t = 4.5$ Gyr, at the end of the first phase of buckling. The white arrows indicate the bulk motion of the
stars in a given direction, and the small black arrows on the orbit (black line) indicate the direction of motion of the
star along the orbit.}
\label{xzvelmaps}
\end{figure}

Interestingly, plotting $f_z'$ (after buckling) as a function of $f_x$ (before buckling), one still obtains
approximately the 2:1 resonance (middle panel of Fig.~\ref{freqzx}). This means that on the way to the resonance only
$f_z$ changes, while $f_x$ remains approximately constant; if the resonance is present, even if $f_x$ is not yet
changed it means that only the variation of $f_z$ is responsible for getting to the resonance. It does not mean that
$f_x$ remains constant during the whole evolution, only that it changes mostly after the resonance is reached. This
observation suggests a simple scenario, in which the buckling event can be divided into two distinct phases: the
first one, where $f_z$ decreases while $f_x$ remains approximately constant; and the second one, where $f_x$ increases
while $f_z$ remains unchanged. The next two sections describe the two phases in more detail.

Before moving on, it may be interesting to look at the evolution of the frequencies once again, but this time in the
form of histograms of the ratio $f_z/f_x$, often discussed in similar studies \citep{Portail2015}.
Figure~\ref{frequencyratios} shows how the ratio evolved, from $f_z/f_x > 2$ before buckling (blue), to $f_z'/f_x \sim
2$ during buckling (green), and $f_z'/f_x' \le 2$ after buckling (red). I remind the reader that the green histogram
shows a combination of the frequencies measured before buckling ($f_x$) and after buckling ($f_z'$), rather than real
frequency estimates during buckling; however, the distribution, although rather wide, still peaks very close to the
value of two expected for the vertical resonance.

\section{First phase: Growing the distortion}

Figure~\ref{distortion1} shows the distortion of the stellar component of the galaxy along the vertical direction ($z$)
as it develops in the first stage of buckling. The maps in the subsequent panels show the mean position of the stars
along $z$ in different simulation outputs separated by $0.05$ Gyr, starting with $t = 4.1$ Gyr. In
the following panels, the distortion becomes more and more pronounced, preserving its shape with respect to the bar,
which is oriented along the $x$-axis in all panels, until about $t = 4.5$ Gyr, after which time it starts to wind up.

The initial stage of the formation of the distortion can be described using the Mathieu equation
\citep{Binney1981}. This formalism predicts instability of stellar orbits with resonant frequencies of $f_z = 2 f_x$,
which are thus able to increase their oscillations in the $z$ direction. However, as can be seen from
Fig.~\ref{freqzx} (left panel) and Fig. 14 of L19 (middle left panel), the number of stars with such
frequencies is very low before buckling. Although weakly populated, the resonance is by no means narrow, since the
stars with resonant frequencies lie in a wide range of radii: between 0.5 and 4 kpc. The mechanism that is able to
increase the number of such stars can be described in a simple way in terms of a driven harmonic oscillator.

I now suppose that some stellar orbits of the bar are caught at vertical resonance and that a small distortion out of
the disk plane forms in the bar, such as the one shown in the upper left panel of Fig.~\ref{distortion1}. The other
stars will feel a vertical force with a distribution tracing that of the existing distortion. The vertical resonance
thus provides a seeding perturbation, which modifies the first orbits, and they change the mass distribution in the bar
and its potential. The modified potential then changes the shape of subsequent orbits, which increase the overall
distortion creating a kind of feedback loop. The picture involving the forced harmonic oscillator also helps us
understand why one type of distortion (smile or frown) is selected and maintained. Once a small distortion appears due
to the resonance, it builds up and enhances the same shape, because the force from the distorted mass distribution can
only modify the orbits into this particular shape.

The vertical oscillations of the stars can therefore be described by the modified equation of the driven
harmonic oscillator of the form \begin{equation}        \label{drivenoscillator}
        \ddot{z} + \omega^2 z = F_0 (1 - \rm{e}^{- \alpha t}) \cos (\omega_{\rm f} t),
\end{equation}
where $\omega$ is the oscillation frequency of the star in the vertical direction, $\omega_{\rm f}$ is the frequency
of the force, while $F_0$ and $\alpha$ are constants. The classical term $\cos (\omega_{\rm f} t)$ describes the
variability of the force due to the distortion of the mass distribution. The additional term $(1 - \rm{e}^{-
\alpha t})$ represents its growth in time up to the maximum value, $F_0$, with the growth rate controlled by $\alpha$.

Let us express the vertical frequencies in terms of the horizontal frequencies, $\omega = f_z/f_x$. As argued in the
previous section, we can assume that the horizontal frequencies, $f_x,$ do not change in this stage of buckling. From
the previous studies (L19), we know that before buckling, $2 < f_z/f_x < 4$; the distribution of these frequency ratios
is shown in Fig.~\ref{frequencyratios} in the form of the blue histogram. Expressed in the same way, the force
frequency $\omega_{\rm f} = 2$.

\begin{figure*}
\centering
\includegraphics[width=6cm]{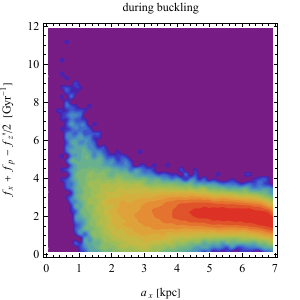}
\includegraphics[width=6cm]{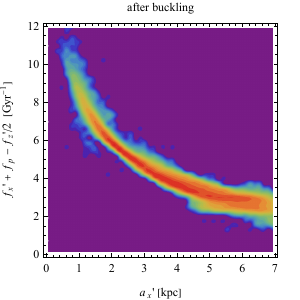}
\includegraphics[width=6cm]{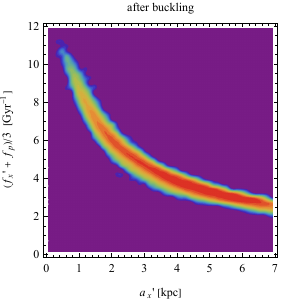}\\
\caption{
Pattern speed of vertical distortion during and after buckling as function of amplitude of oscillations
along the bar. The left panel shows the approximately constant pattern speed during buckling, assuming the evolved
$f_z'$ and unevolved $f_x$. The middle panel shows the strongly decreasing pattern speed after buckling with evolved
$f_z'$ and $f_x'$. The right panel shows $(f_x' + f_{\rm p})/3,$ which turns out to be equal to the pattern speed
after buckling.}
\label{freqafter}
\end{figure*}

The solution of Equation~(\ref{drivenoscillator}) with the natural choice of initial conditions $z(0) = z_0$ and
$\dot{z} (0) = 0$ can be found analytically with {\em Mathematica} software, but it is too complicated to cite it here
in full. Instead, I discuss a few examples. The well-known outcome of the driven harmonic
oscillator is to change the initial frequency, $\omega,$ to the frequency of the force, $\omega_{\rm f}$, and it is
also the case here. Figure~\ref{resonance} shows three representative orbits that illustrate this transformation. The
upper panels of the figure present the solutions for $z(t)$ for $\omega = 2.5$, 3, and 3.5, which cover the range of
interesting values of $\omega$ found to be present in the bar before buckling (Fig.~\ref{frequencyratios}). For all
orbits, the values of the constants were assumed to be $F_0 = 10$ and $\alpha = 0.1$, which turn out to produce a
sufficient effect.

In each panel, the blue line corresponds to the initial (unforced) oscillation of the star $z(t) = z_0 \cos(\omega t),$
and the red one to the solution under the effect of the force. One can see that the effect of the force is to increase
the amplitude of the oscillation (from the initial small value of $z_0 =0.5$ kpc), but also to change its frequency.
While within $4 \pi$ the star oscillates five, six, and seven times for $\omega = 2.5, 3,$ and 3.5, respectively,
before the forcing, it only oscillates four times after forcing. It should be noted that the orbits with initial
frequencies differing more from the frequency of the force need more time to transform.

The lower panels of Fig.~\ref{resonance} show the corresponding 3D orbits of the stars in the form of the Lissajous
curves combining the motion along $z$ with the ellipse in the $xy$ plane. The ellipses are the x1-type orbits, which
are well known from classical studies of the orbital structure of bars \citep{Contopoulos1980, Skokos2002}. The blue
and red orbits of the different columns show the 3D motion of the stars without and with the action of the force, and
the three panels in the lowest row show them together for comparison. While the modified orbits are not exactly the
banana-type orbits because of the variation of the amplitude, they are remarkably similar.

One can now imagine how the stellar orbits in the bar evolve in this phase of buckling. After a small perturbation out
of the disk plane appears, the stars with $\omega$ only slightly larger than two will have their amplitudes
strongly increased (left column of Fig.~\ref{resonance}), thus increasing the force. This in turn will affect orbits
with higher values of $\omega,$ bringing them all to $\omega \sim 2$ (middle and right columns of Fig.~\ref{resonance}).
This is what we see in the first nine panels of Fig.~\ref{distortion1}: the distortion grows as more orbits join the
resonance preserving its orientation with respect to the bar. This process takes about 0.4 Gyr (from $t = 4.1$ till $t
= 4.5$ Gyr) in this simulation. At $t = 4.5$ Gyr, many of the stars that were on x1 orbits initially now follow the
banana-like resonant orbits.

This simple description makes it possible to understand how the distortion can build up, creating orbits that are
similar in shape to the well-known banana orbits. Such orbits would produce a distortion that is symmetrical with
respect to the $x$-axis of the bar. As one can see from Fig.~\ref{distortion1}, this is not exactly the case. While
the yellow and red part of the distortion, signifying departure above the disk plane, is approximately aligned with
the bar ends, the blue part (below the plane) is not exactly perpendicular to the bar. This means that the
average orbit of the stars at this stage looks more like the one shown in Fig.~\ref{resorbit}, which has its minima
shifted with respect to the $y$-axis.

Interestingly, this kind of orbit also explains the mean velocity of stars during this phase of buckling discussed
by \citet{Li2023}. Figure~\ref{xzvelmaps} shows the mean velocity of the stars in the horizontal and vertical directions,
$v_x$ and $v_z$ (upper and lower panels, respectively) at the time of maximum distortion ($t = 4.5$ Gyr). The motion of
the stars can be understood if on average the stars follow the orbit shown in black. The components of the stellar
motion (black arrows along the orbit) along the $x$ and $z$ directions agree with the overall average motion of all
stars in these directions shown by the white arrows. This average orbit probably originates from the combination of
banana and anti-banana ($\infty$-like) orbits.

The maximum distortion of the bar in the first phase of buckling, at time $t_{\rm b} = 4.5$ Gyr, can be approximated as
$z_{\rm d} (R, \phi, t_{\rm b}) = z_0(R) \cos m \phi$ with $m=2$. As discussed by \citet{Binney2008} in the context of
galactic warps (their Section 6.6.1), the evolution of such a distortion can be described as a propagation of a
kinematic bending wave with a pattern speed of $\omega_{\rm p} = \Omega - \nu/2$ (a different symbol, $\omega_{\rm p},$
is used here to distinguish it from the pattern speed of the bar, $\Omega_{\rm p},$ or $f_{\rm p}$ in the notation of
this paper). At resonance during buckling and using the notation adopted here, $\omega_{\rm p} = f_x + f_{\rm p} -
f_z'/2 = f_{\rm p,}$ since $f_z' = 2 f_x$ at this time. This means that the bending wave propagates with the pattern
speed of the bar, and therefore the distortion pattern seen in Fig.~\ref{distortion1} remains stationary in the
reference frame of the bar until $t_{\rm b} = 4.5$ Gyr. This is confirmed by the distribution of $f_x + f_{\rm p} -
f_z'/2, $ which is plotted as a function of the amplitude of the oscillations along the horizontal axis, $a_x$, in the
left panel of Fig.~\ref{freqafter}. This combination of frequencies remains approximately constant with radius during
buckling. Here, as before, it is assumed that the frequencies along the vertical direction have already changed
($f_z'$), while those along the bar have not yet evolved ($f_x$).

\begin{figure}
\centering
\includegraphics[width=4.6cm]{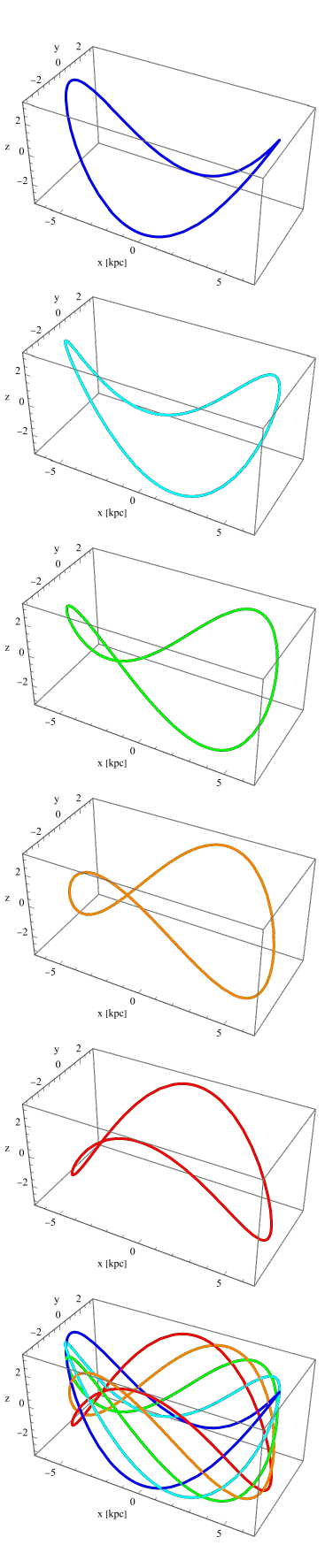}\\
\caption{
Passage of bending wave. The images from top to bottom illustrate the evolution of the distortion in time at one
selected distance from the center of the bar. As the minimum of the curve moves along the ellipse in the $xy$ plane,
the distortion changes from smile-like to frown-like in the edge-on view. The lowest image shows the
combination of all the shapes.}
\label{wave}
\end{figure}

\begin{figure}
\centering
\includegraphics[width=8cm]{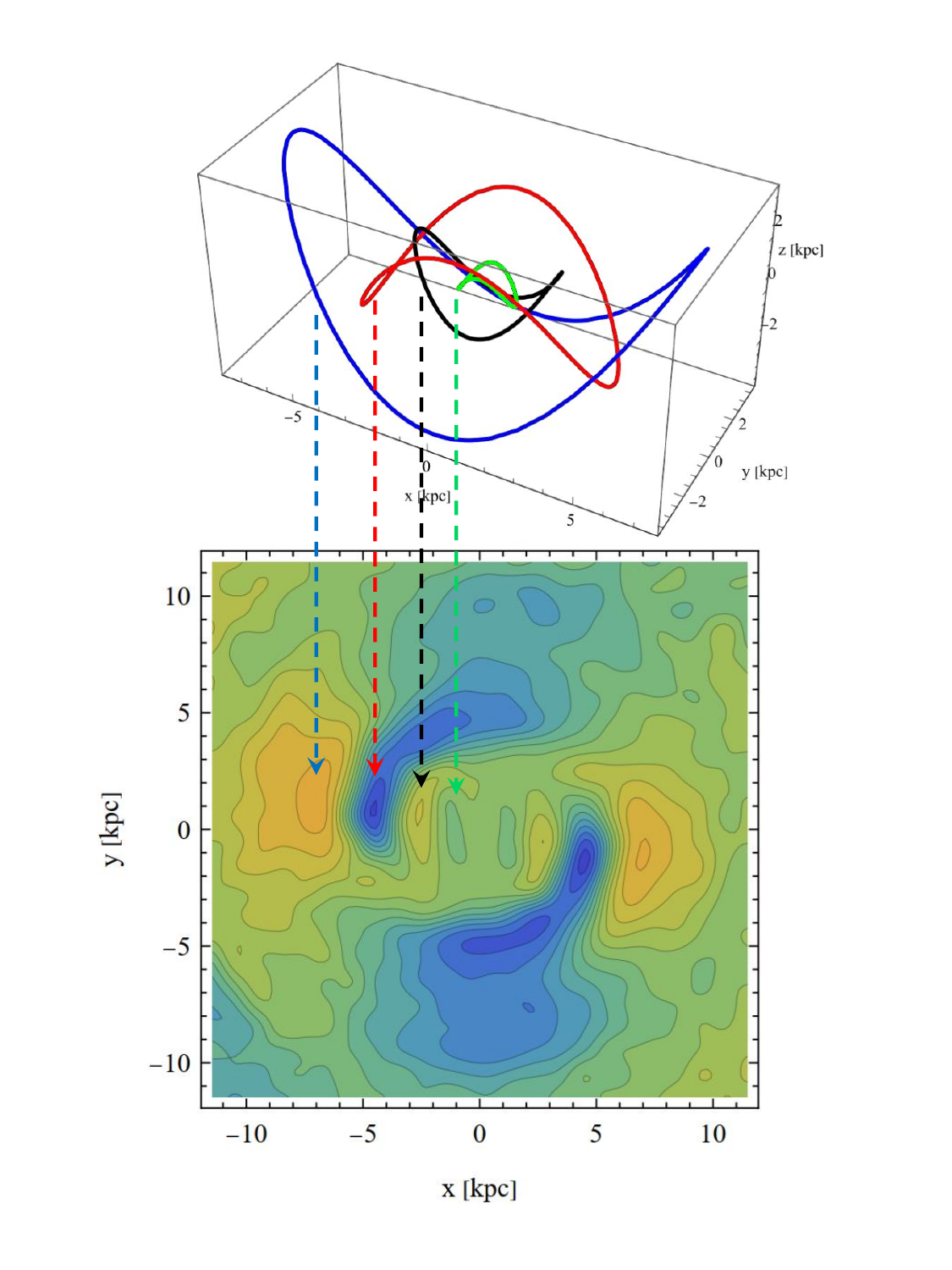}\\
\vspace{-0.5cm}
\includegraphics[width=8.1cm]{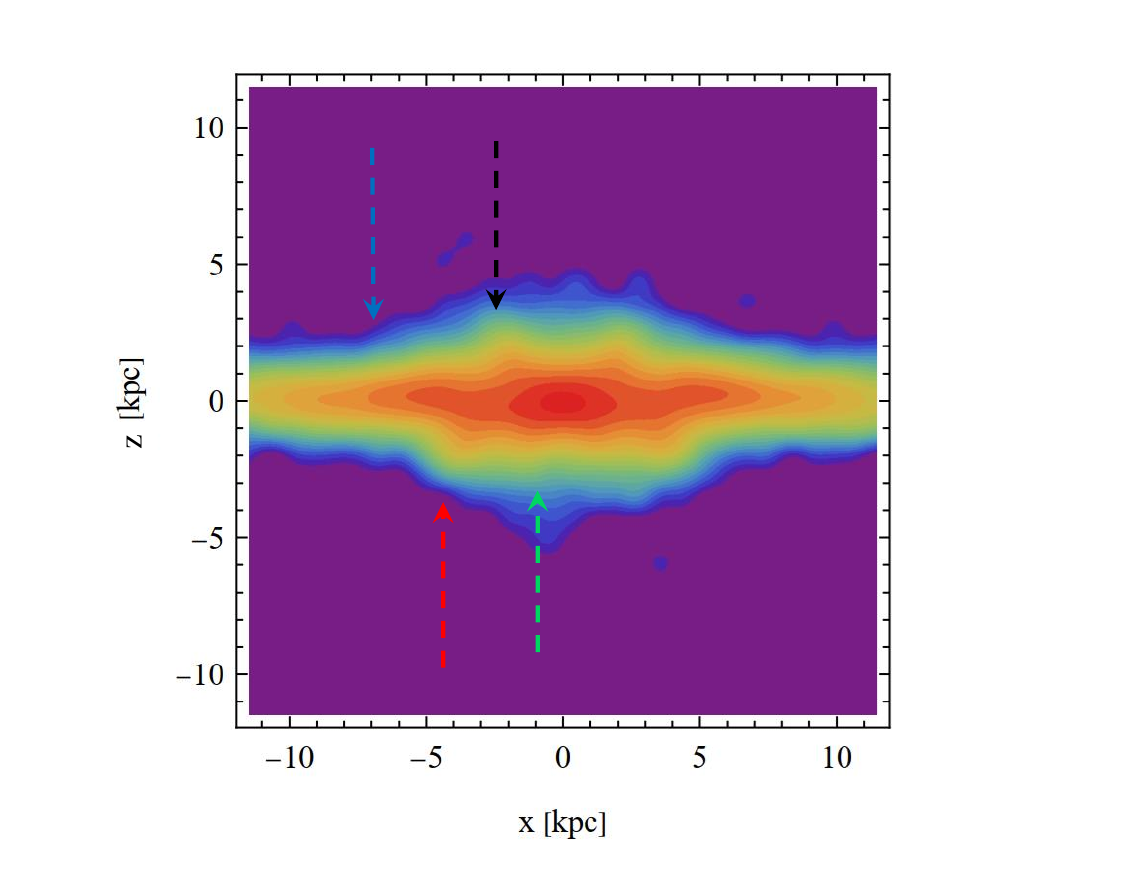}\\
\caption{
Configuration leading to distortion seen in map of average positions of stars along vertical
direction (middle panel) and in surface density of stellar component seen edge-on (along $y$, lower panel) for
one selected time, $t = 4.85$ Gyr. The dashed arrows indicate the correspondence between the distortions rotated by
the bending wave differently at different radii and the maxima and minima of the distributions of the stars in the
face-on position map and edge-on density map.}
\label{configuration}
\end{figure}

\begin{figure*}
\centering
\includegraphics[width=7.5cm]{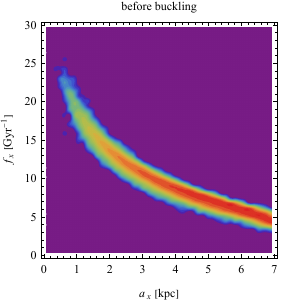}
\includegraphics[width=7.5cm]{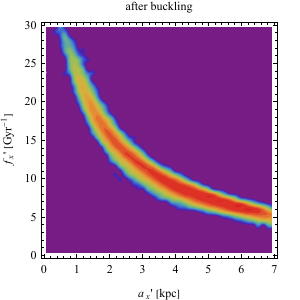}\\
\caption{
Dependence of frequency of oscillations along bar on their amplitude before (left panel) and after (right panel)
buckling.}
\label{fxax}
\end{figure*}

\begin{figure*}
\centering
\includegraphics[width=15cm]{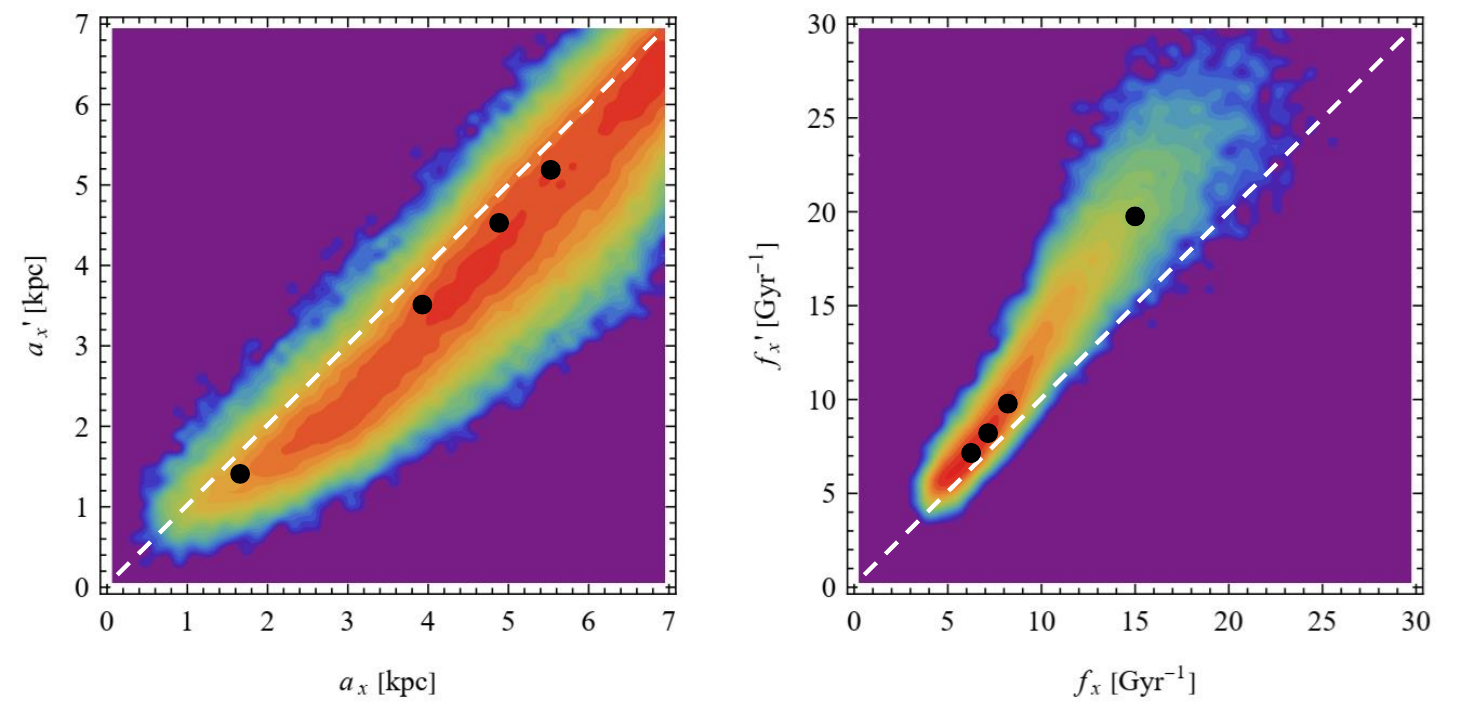}\\
\caption{
Relations between amplitudes of oscillations along bar (left panel) and their frequencies (right panel) before and
after buckling. White dashed lines indicate equality between the measured values and the black points mark the
values corresponding to the first four orbits shown in Fig.~\ref{pretzels}, with properties listed in the first four
rows of Table~\ref{properties}.}
\label{fxaxadiabatic}
\end{figure*}

\begin{table*}
\caption{Properties of selected stellar orbits before and after buckling.}
\label{properties}
\centering
\begin{tabular}{c c c c c c c c}
\hline\hline
$f_x'$ [Gyr$^{-1}$] & $f_z'$ [Gyr$^{-1}$]& $f_z'/f_x'$ & $f_x$ [Gyr$^{-1}$]&$a_x'$ [kpc]&$a_x$ [kpc]& $\psi$& $n$     \\
\hline
20                  &  30                & 3/2         & 15                & 1.4        & 1.6       & 1/3   & 2       \\
10                  &  16.67             & 5/3         & 8.33              & 3.5        & 3.8       & 1/5   & 3       \\
8                   &  14                & 7/4         & 7                 & 4.6        & 4.9       & 1/7   & 4       \\
7.14                &  12.86             & 9/5         & 6.43              & 5.2        & 5.5       & 1/9   & 5       \\
5                   &  10                & 2           & 5                 & 7.3        & 7.3       & 0     & $\infty$\\
\hline
\end{tabular}
\end{table*}

\section{Second phase: Winding up the distortion}

The evolution of the distortion in the vertical direction changes after $t_{\rm b} = 4.5$ Gyr. As demonstrated by
the three panels of the lowest row in Fig.~\ref{distortion1} ($t = 4.55 - 4.65$ Gyr), the pattern starts to wind up and
continues to do so in the later stages illustrated in Fig.~\ref{distortion2} of the appendix. The passage of the
bending wave at one particular distance from the center of the bar is visualized in Fig.~\ref{wave}, where the
distortion of the bar is seen to evolve from smile-like to frown-like in the edge-on view. The same happens at
all radii, but at a different rate, which leads to the more complicated distortion patterns seen at later times in
Fig.~\ref{distortion2}.

The connection between the passage of the bending wave and the distortion patterns is illustrated in
Fig.~\ref{configuration} for $t = 4.85$ Gyr, one of the outputs shown in Fig.~\ref{distortion2}. The  maximum
distortion was smile-like at all radii at $t_{\rm b} = 4.5$ Gyr. Between $t = 4.5$ and 4.85 Gyr, the bending wave
rotated the distortion differently at different radii, and more so for those closer to the center (upper panel).
This evolution creates a pattern of maxima and minima in the map of the average positions of the stars (middle panel)
along the bar. The same pattern can also be discerned in the edge-on view of the surface density of the stellar
component of the galaxy at this time (lower panel).

Figure~\ref{distortion2} shows that as the winding-up progresses, the pattern of maximal and minimal distortion moves
outward toward the end of the bar but is still clearly visible, for example at $t = 5.05$ Gyr. Later on the distortion
becomes so tightly wound that it ceases to be visible in the inner 5 kpc. The distortion, however, is still present at
radii of $R \sim 7$ kpc, so in a sense buckling continues at these outer radii, as discussed below.

Thus, in the second phase of buckling the pattern speed of the bending wave is no longer constant with radius. Indeed,
the middle panel of Fig.~\ref{freqafter} shows that after buckling $f_x' + f_{\rm p} - f_z'/2$ decreases strongly as a
function of $a_x'$ (now all the quantities are primed and measured in the period of time after buckling, as discussed
in Section~2). Interestingly, this combination of frequencies turns out to behave almost exactly the same way as $(f_x'
+ f_{\rm p})/3$ (or $\Omega'/3$ in the standard notation), which is shown in the right panel of Fig.~\ref{freqafter}.
The only difference is that in the middle panel a small remnant of the constant pattern speed remains for 4.5 kpc $<
a_x' <$ 7 kpc. This corresponds to the small peak of the remaining resonant (banana) orbits with $f_z'/f_x' = 2$ seen
in the red histogram of Fig.~\ref{frequencyratios}. The equality between these two expressions, $f_x' + f_{\rm p} -
f_z'/2 = (f_x' + f_{\rm p})/3,$  immediately leads to Equation~(\ref{relationfxfz}) after rearrangement of the terms.

It should be noted that a similar change in the behavior of $f_x' + f_{\rm p} - f_z'/2$ after buckling can also be
seen in Fig.~13 of \citet{Li2023}. After buckling (at $t = 3.6$ Gyr in their simulation) the same quantity, $\Omega -
\nu_z/2$ in their notation (blue curve), increases and traces $\Omega$ (black curve) so that approximately $\Omega -
\nu_z/2 = \Omega/3$, in agreement with the finding presented here.

The change in the behavior of the pattern speed of the bending wave after buckling can be traced to the change in the
frequency along the bar, since this is the only quantity that varies from $f_x$ to $f_x'$ between the left and middle
panels of Fig.~\ref{freqafter}. Indeed, by comparing $f_x (a_x)$ before buckling with $f_x' (a_x')$ after buckling in
Fig.~\ref{fxax}, one can immediately see that the horizontal frequencies increase. This can be further appreciated in
Fig.~\ref{fxaxadiabatic}, which shows separately how the amplitude of the oscillations and the frequency evolved. This
figure demonstrates that after buckling the amplitudes were on average decreased ($a_x' < a_x$), while the frequencies
were increased ($f_x' > f_x$). The difference between the frequencies before and after buckling vanishes near the end
of the bar at around $R \sim 7$ kpc. In this region, $f_x' + f_{\rm p} - f_z'/2 = (f_x' + f_{\rm p})/3$ is still equal
to the pattern speed of the bar $f_{\rm p}$ because $f_x' = 2 f_{\rm p}$, and this is why the distortion survives.

\section{Making pretzels from bananas}

I now consider a few examples of horizontal frequencies of stellar orbits after buckling, $f_x'$, from the range shown
in the right panel of Fig.~\ref{fxax}. The selected values are listed in the first column of Table~\ref{properties}.
Using Equation~(\ref{relationfxfz}), one can immediately find the corresponding vertical frequencies after buckling,
$f_z'$. These values are given in the second column of the table, and their ratios, $f_z'/f_x'$, are given in the
third. The range of values of these ratios, which is between 1.5 and 2, covers the whole set seen in the red histogram
in Fig.~\ref{frequencyratios}. Given that during buckling $f_z' = 2 f_x,$ one can immediately recover the initial
horizontal frequencies, $f_x$ (fourth column). The typical amplitudes of the oscillations along the bar after and
before buckling, giving an estimate of the size of the stellar orbits in this dimension, $a_x'$ and $a_x$, can then be
found from fitting the very tight relations $f_x' (a_x')$ and $f_x (a_x)$ in Fig.~\ref{fxax}. Their values are given in
the fifth and sixth columns of Table~\ref{properties}.

\begin{figure}
\centering
\vspace{-1.5cm}
\includegraphics[width=7.7cm]{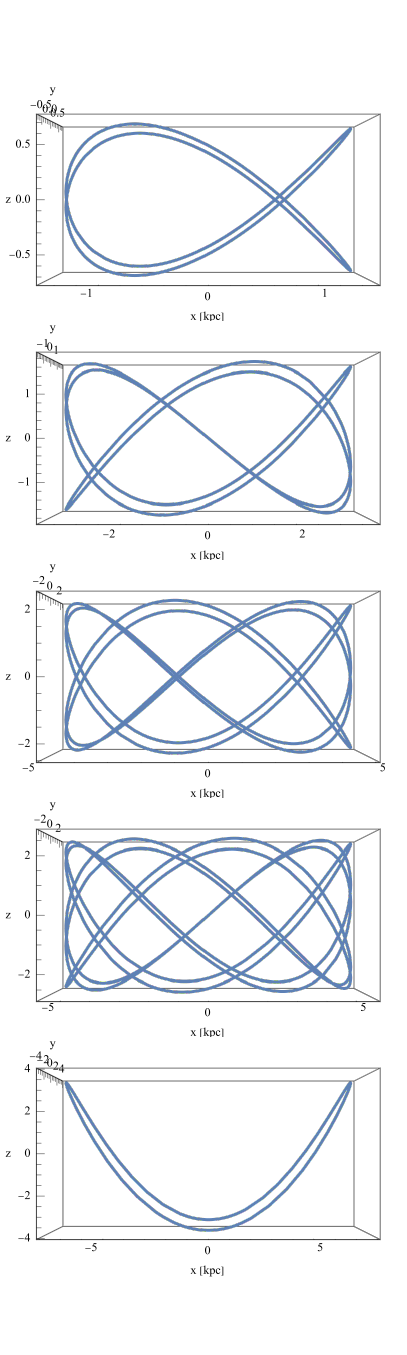}\\
\vspace{-1cm}
\caption{
Examples of pretzel orbits formed as result of buckling, viewed edge-on. The banana orbit is included in the lower
panel. The frequency ratios of the orbits are $f_z'/f_x' =3/2$, 5/3, 7/4, 9/5, and 2, from top to bottom.}
\label{pretzels}
\end{figure}

The passage of the bending wave may be interpreted as speeding up the motion of a star in the horizontal plane. For
example (all frequencies in units of 1 Gyr$^{-1}$), a star with $f_x=7$ made seven cycles in 1 Gyr, and after buckling
it makes eight due to the bending wave: $f_x' = f_x + (1/7) f_x$; a star with $f_x=25/3=8.33$ now has $f_x' = 25/3 +
(1/5) 25/3 = 10$, while a star with $f_x=15$ makes five more cycles and has $f_x'= f_x +(1/3) f_x = 20$. These
regularities can be generalized to
\begin{equation}        \label{fxpsi}
        f_x' = f_x (1 + \psi),
\end{equation}
with the values of $\psi$ given in the penultimate column of Table~\ref{properties}. The dependence of $\psi$ on the
horizontal frequency can be approximated as
\begin{equation}        \label{psi}
        \psi =  (f_x - 2 f_{\rm p})/(2 f_x) = (f_x' - 2 f_{\rm p})/(2 f_x' + 2 f_{\rm p}),
\end{equation}
where $f_{\rm p}$, as always, is the pattern speed of the bar. The formula makes the special significance of $f_x
= f_x' = 2 f_{\rm p}$ clear, because $\psi = 0$ for this particular frequency. It means that the stellar orbits with
frequencies equal to twice the pattern speed of the bar are not modified after buckling (last row of
Table~\ref{properties}).

The transformation of other stellar orbits after buckling can be understood using 3D Lissajous curves. These
curves can be parametrized as
\begin{equation}        \label{lissa}
        (x, y, z) = [a \cos \phi, \ b \sin \phi, \ c \cos (m \phi + \phi_0)],
\end{equation}
where $a,b,c$ are the amplitudes of the oscillations along $x,y,z$, respectively, and $m$ is the ratio of
the vertical and the horizontal frequency. It is assumed here that in the horizontal plane $xy$ the orbits were, and
remain, ellipses; that is, they originate from the x1 orbits. In this case, $f_y = f_x$ before buckling and $f_y' =
f_x'$ after buckling. On the other hand, the ratio of the vertical and the horizontal frequency varies during buckling
and changes from $f_z/f_x > 2$ before the event to $f_z'/f_x' \le 2$ afterwards, as discussed above. For resonant
orbits, $f_z' = 2 f_x,$ so a banana orbit is obtained for $m=2$ in Equation~(\ref{lissa}). With the constant $\phi_0 =
0$, varying $\phi$ from 0 to $2 \pi,$ we obtain the ``pure'' banana shown in the lower panel of Fig.~\ref{pretzels}.
For this plot, it was assumed that $a = a_x' = 7.3$ kpc (in agreement with the values from Table~\ref{properties} for
$f_x = f_x' =5$ Gyr$^{-1}$) and $b = c = a/2$. Adopting a nonzero value of $\phi_0$ shifts the banana orbit toward an
$\infty$-like orbit; for example, the orbit shown in Fig.~\ref{resorbit} was obtained with $\phi_0 = -\pi/5$.

At the end of the first phase of buckling, all orbits can be approximated as banana-like orbits with $f_z' = 2
f_x$. Using Equation~(\ref{fxpsi}), one obtains a relation between the frequencies after buckling:
\begin{equation}        \label{pre}
        f_z' = 2 f_x'/(1 + \psi) = f_x' (2 - 1/n),
\end{equation}
where $n = (1 + \psi)/(2 \psi)$ and the values of the parameter $n$ for the selected orbits are given in the last
column of Table~\ref{properties}. In order to obtain the new orbits from the Lissajous curves given in the parametric
form (\ref{lissa}), one needs to replace $m=2,$ which is valid for the banana orbits, with $(2 - 1/n)$. These values
are obviously equal to the $f_z'/f_x'$ ratios given in the third column of Table~\ref{properties}. The orbits obtained
in this way using the Lissajous curves are shown in Fig.~\ref{pretzels}. The sizes of the orbits along $x$ were
adjusted to the values of $a = a_x'$ from Table~\ref{properties}, while $b = c = a/2$ and $\phi_0 = 0$ were kept for
all of them. It should be noted that in order to close, the orbits are required to cover $4 \pi$, $6 \pi$, $8 \pi$, and
$10 \pi$ in $\phi$ for $n = 2,3,4,$ and 5, respectively.

The origin of these pretzel-like orbits can be understood by recalling that the bending wave speeds up the motion in
the horizontal plane. I now focus on the first example of a fish-like orbit shown in the upper panel of
Fig.~\ref{pretzels}. Starting at the upper right corner of the box, at the highest $z$ position and following the
orbit, one can see that in comparison with the original banana orbit (bottom panel) the star indeed reaches the next
maximum of $z$ not after half a cycle, but much further along the horizontal angle. Overall, the result is that the
star makes only three full oscillations along $z$ for each two oscillations in $x$, as expected from the frequency
ratio for this orbit, $f_z'/f_x' =3/2$. Similar behavior is observed for the next orbits shown in Fig.~\ref{pretzels},
but to a lesser extent. The orbits with $f_z'/f_x' =$ 5/3, 7/4, and 9/5 make five, seven, and nine oscillations in $z$
for each three, four, and five oscillations in $x$, respectively. The limiting case of this sequence is obviously the
banana orbit with $f_z'/f_x' = 2$, which remains unchanged.

The positions of the first four orbits from Table~\ref{properties} and Fig.~\ref{pretzels} in the $a_x' (a_x)$ and
$f_x' (f_x)$ planes are shown as black dots in Fig.~\ref{fxaxadiabatic}. The variation of these two quantities in the
second phase of buckling suggests that they change adiabatically. As discussed in Section 3.6.2 of \citet{Binney2008},
the increase of the horizontal frequency of a harmonic oscillator by a factor $(1 + \psi)$ described by
Equation~(\ref{fxpsi}) should correspond to the decrease of the amplitude of oscillations according to $a_x' = a_x (1 +
\psi)^{-1/2}$. The numerical values in Table~\ref{properties} indeed obey these relations.

\section{Discussion}

In this work, I revisited the issue of the origin and interpretation of the buckling instability in galactic bars. It
was proposed that the phenomenon can be divided into two distinct phases. The first one involves the growth of the
distortion of the bar out of the disk plane due to the vertical resonance of stellar orbits in the bar. The growth of
the distortion can be described with the mechanism of the driven harmonic oscillator where the initial small
perturbation acts as a driving force and modifies the vertical frequencies of other orbits. The distortion retains its
orientation with respect to the bar until reaching a maximum value.

The increase of the amplitudes of the oscillations of the stars in the vertical direction is accompanied by a decrease
of the amplitudes along the bar and the bar shrinks leading to the increase of the mass in the central parts, as
noticed by \citet{Li2023}. The harmonic oscillators then seem to respond adiabatically, increasing the frequencies of
the oscillations along the bar. Indeed, the factors by which the horizontal frequencies and amplitudes change agree
with this interpretation.

As a result of the increased horizontal frequencies, in the second phase of buckling the pattern speed of the vertical
distortion is no longer equal to the pattern speed of the bar -- it is higher -- and the distortion starts to wind up.
The banana orbits that dominated the orbital structure of the bar at the time of the resonance evolve into pretzel-like
orbits with vertical-to-horizontal frequency ratios below two. The winding up of the distortion also explains why the
bar becomes weaker after buckling. Namely, the winding up is associated with faster rotation of the stars in the
horizontal plane, which means that the orbits become less radial.

The equation of a driven harmonic oscillator (although without the force growing in time) was previously used in the
context of bar buckling by \citet{Collier2020} in an attempt to relate buckling to the phenomenon of Landau damping.
Here, instead, this equation was used to demonstrate how the initial small deviation from the vertical symmetry can
build up, leading to a strong global distortion of the bar, and the scenario bears no relation to Landau damping.

The violent buckling event seen in the simulation described here, leading to the strong changes in the orbital
structure of the bar, seems to be characteristic of pure disks embedded in dark-matter halos, with no bulge component.
In a recent study, \citet{McClure2025} looked into the phenomenon using simulations with different bulges of increasing
mass. In the case without the bulge, their results are very similar to the ones discussed here; namely, after buckling
the orbital frequencies display a very tight relation similar to the one shown in the right panel of Fig.~\ref{freqzx}.
With increasing bulge mass, the resonant orbits still appear, but they do not evolve strongly toward
vertical-to-horizontal frequency ratios below two, and they remain close to this resonant value. It seems, therefore,
that the bulge does not inhibit the appearance of the vertical resonance, but it does prevent the passage to the second
phase of buckling as described here. This is understandable given the importance of mass redistribution in changing the
horizontal frequencies, mentioned above. With the additional massive component in the center, this must certainly look
different.

Recently, \citet{Zozulia2024} studied the vertical evolution of a mature simulated bar, long after the violent buckling
event. They reported behavior similar to that described here, namely the catching of the orbits at resonance and their
later passage to vertical-to-horizontal frequency ratios below two, which they refer to as the ``circulation mode''.
This stage of bar evolution corresponds to the period after $t = 5$ Gyr in the simulation discussed here when the
distortion is preserved only in the outer part of the bar (last panels of Fig.~\ref{distortion2}). As discussed in L19,
this distortion survives for a few gigayears, shifting toward the outer part of the bar and seeding the second episode
of buckling later on. It therefore seems that as the bar grows, its outer parts continuously increase amplitudes of
their oscillations in the vertical direction in a kind of quiescent buckling. The process described by
\citet{Zozulia2024} is thus not essentially different from the violent buckling event studied here. The latter only
happens on a shorter timescale and in the whole inner part of the bar. This picture of the moving distortion also
agrees with the early results of \citet{Pfenniger1991}, which found (their Fig. 14) that the position of the vertical
resonance shifts to larger radii after the formation of the BP shape.

A few open questions remain concerning the analysis presented here. For example, it is still not clear why the pattern
speed of the vertical distortion increases in such a particular way in the second stage of buckling and becomes equal
to a third of the circular frequency. A related question concerns why the banana orbits in the outer part of the bar
remain unchanged. This region is characterized by the horizontal frequency equal to twice the pattern speed of the bar,
so the pattern speed of the vertical distortion remains equal to the pattern speed of the bar and the distortion does
not wind up there.

In summary, this work demonstrates that the phenomenon of buckling instability is not related to the fire-hose
instability known from plasma physics, but instead it is triggered by the vertical resonance of bar orbits. The resonance
leads to the growing distortion of the bar, which later winds up, resulting in a transformation of banana-like orbits
into pretzel-like ones and the formation of a BP shape.

\begin{acknowledgements}
I am grateful to the anonymous referee for useful comments. Computations for this work have been performed
using the computer cluster at the Nicolaus Copernicus Astronomical Center of the Polish Academy of Sciences (CAMK PAN).
\end{acknowledgements}

\begin{appendix}

\onecolumn

\section{Later evolution of the distortion pattern}

This appendix presents the continuation of Fig.~\ref{distortion1} and shows the evolution of the vertical distortion
pattern of the bar after it starts to wind up.

\begin{figure*}[h!]
\centering
\includegraphics[width=5cm]{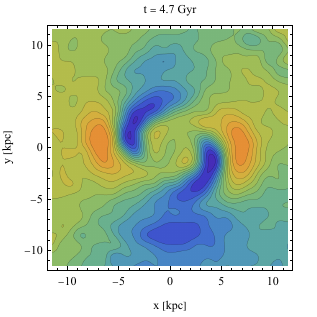}
\includegraphics[width=5cm]{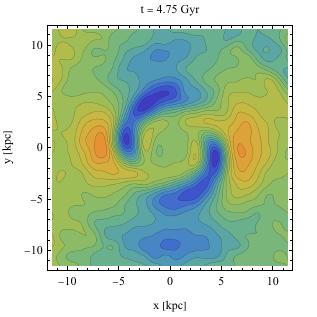}
\includegraphics[width=5cm]{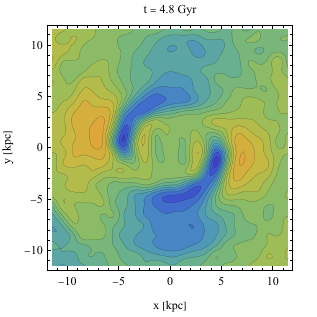}\\
\vspace{0.15cm}
\includegraphics[width=5cm]{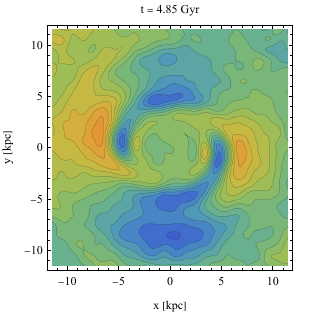}
\includegraphics[width=5cm]{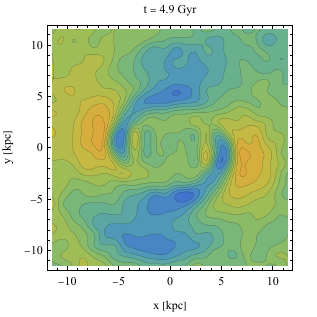}
\includegraphics[width=5cm]{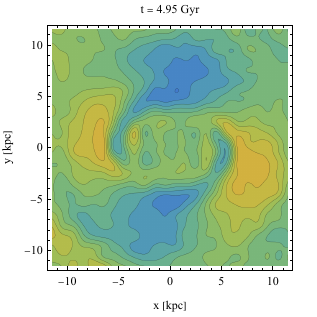}\\
\vspace{0.15cm}
\includegraphics[width=5cm]{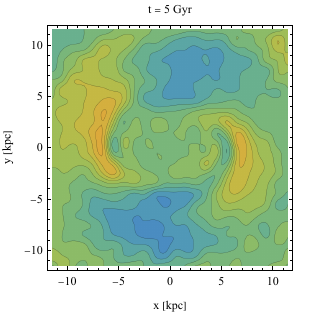}
\includegraphics[width=5cm]{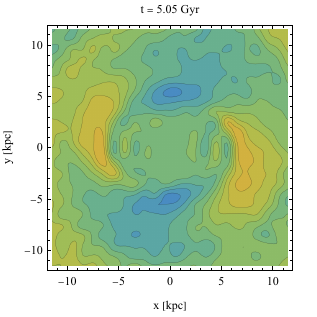}
\includegraphics[width=5cm]{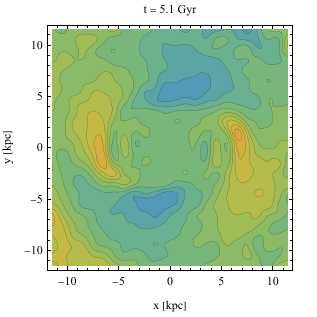}\\
\vspace{0.15cm}
\includegraphics[width=5cm]{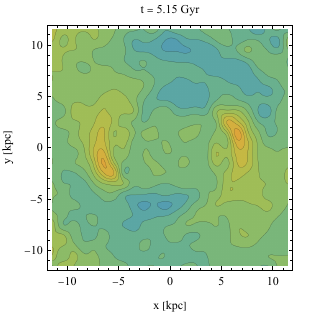}
\includegraphics[width=5cm]{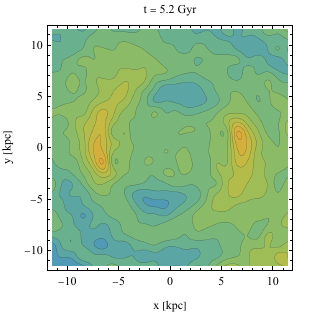}
\includegraphics[width=5cm]{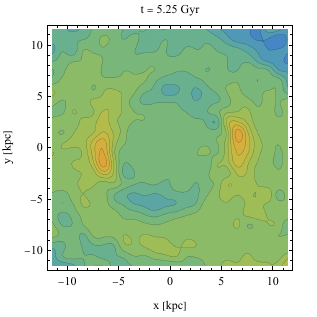}\\
\vspace{0.15cm}
\hspace{0.25cm}
\includegraphics[width=4.06cm]{legend14.pdf}\\
\caption{
Same as Fig.~\ref{distortion1}, but for later times, $t = 4.7$ to 5.25 Gyr.}
\label{distortion2}
\end{figure*}

\end{appendix}

\end{document}